\documentclass[amsmath,amssymb,aps,floats,prl,twocolumn,showpacs,unsortedaddress,superscriptaddress]{revtex4}
\usepackage{graphicx}
\usepackage{amssymb}
\usepackage{epstopdf}
\usepackage{color}
\usepackage[normalem]{ulem}

\newcommand{\ie}{\emph{i.e.} }

\newcommand{\ohm}{~\Omega }

\newcommand{\be}{\begin{eqnarray}}
\newcommand{\ee}{\end{eqnarray}}
\newcommand{\bfig}{\begin{figure}}
\newcommand{\efig}{\end{figure}}

\newcommand{\crnbs} {Cr$_{1/3}$NbS$_2 $}
\begin{document}
%\preprint{} 

%\title{{\red{new title needed:}}Spin-Orbit Coupling Induced Anisotropy in the Magnetotransport of the Chiral Helimagnet Cr$_{1/3}$NbS$_2$}%
\title{Out-of-Plane Spin-Orientation Dependent  Magnetotransport Properties in the  Anisotropic Helimagnet \crnbs}

\author{Alexander C. Bornstein}
\affiliation{Department of Physics, University of Colorado, Boulder, CO 80309, USA}%
\author{ Benjamin J. Chapman}
\affiliation{Department of Physics, University of Colorado, Boulder, CO 80309, USA}%
\author{ Nirmal J. Ghimire$^{\dagger}$}
\affiliation{Department of Physics and Astronomy, The University of Tennessee, Knoxville, Tennessee 37996, USA}%
\affiliation{Materials Science and Technology Division, Oak Ridge National Laboratory, Oak Ridge, Tennessee 37831, USA}%
\author{ David G. Mandrus}
\affiliation{Department of Physics and Astronomy, The University of Tennessee, Knoxville, Tennessee 37996, USA}%
\affiliation{Materials Science and Technology Division, Oak Ridge National Laboratory, Oak Ridge, Tennessee 37831, USA}%
\affiliation{Department of Materials Science and Engineering, The University of Tennessee, Knoxville, TN 37996, USA}%
\author{ David S. Parker}
\affiliation{Materials Science and Technology Division, Oak Ridge National Laboratory, Oak Ridge, Tennessee 37831, USA}%
\author{Minhyea Lee}
\email{minhyea.lee@colorado.edu}
\affiliation{Department of Physics, University of Colorado, Boulder, CO 80309, USA}%

\date{\today}

\begin{abstract}
Understanding the role of spin-orbit coupling (SOC) has been crucial to
controlling  magnetic anisotropy in  magnetic multilayer films \cite{ Brooks1940, Bruno1989,Wang1993,Bode2002}.  
It has been shown that electronic structure can be altered via interface SOC by varying the superlattice structure, resulting  in  spontaneous magnetization perpendicular or parallel to the plane ~\cite{Gimbert2012,Hotta2013}.  
In lieu of magnetic thin films, 
we  study the similarly anisotropic helimagnet \crnbs, where  the spin polarization direction, controlled by the applied magnetic field, can modify the electronic structure. As a result, the direction of spin polarization can modulate the density of states, and in turn affect the in-plane electrical conductivity.
In \crnbs, we found an enhancement of in-plane conductivity when the spin polarization is out-of-plane, as compared to in-plane spin polarization.
This is  consistent with  the  increase of density of states  near the Fermi energy at the same spin configuration,  found from first principles calculations.
We also observe unusual field dependence of  the Hall signal in the same temperature range. %of  $T<T_A$, yielding a characteristic temperature scale $T_A$. 
This is unlikely to  originate  from the non-collinear spin texture, but  rather further indicates strong dependence of  electronic structure on spin orientation relative to the plane.
\end{abstract}

\maketitle

Despite the fact that its typical energy scale in $3d$ ferromagnetic metals is small compared to other relevant scales such as  band widths, SOC mixes the nature of the spin and orbital components of the Bloch state in a nontrivial way and leads to a variety of electrical transport phenomena e.g. the anomalous Hall effect (AHE), anisotropic magnetoresistance (AMR), and the planar Hall effect.  
In addition, the recent work  in non-collinear magnetically ordered states  and the related topological Hall effect \cite{Muhlbauer2009,Lee2009,Neubauer2009}  not only has renewed the pivotal role of SOC through the Dzyaloshinskii-Moriya (DM) interaction \cite{Dzyaloshinsky1958,Moriya1960, Hopkinson2009}, but also has  presented a possibility to employ these findings  for functional components in magnetic devices~\cite{Fert2013,Romming2013}.  Non-collinear  magnetic ordering is also suggested to possibly manifest  spin-orbit coupling in a complex manner,  through  the DM interaction\cite{Sales2008, Hopkinson2009, Mokrousov2013}.  Consequently, the modification of  electronic structure by spin-orbit coupling is expected to make in-plane electrical transport  sensitive to the magnetization orientation relative to the plane.

\crnbs~ has a layered crystalline structure, in which 3$d$ transition metal Cr atoms are intercalated in the hexagonal $2H$-type NbS$_2$ matrix as trivalent ions and magnetically order at $T_C =133$ K.  The ferromagnetic layers of Cr$^{3+}$ lie coplanar with the crystallographic $ab$-planes and the magnetic helix propagates along the $c$-axis with a long pitch of 48 nm, corresponding to 40 unit cells \cite{Ghimire2013}.  Its helimagnetic  ordering is attributed to the DM interaction, which originates from a broken inversion symmetry shared by all members of space group $P6_322$ \cite{Moriya1982,Miyadai1983,Ghimire2013}.  With application of a magnetic field ($H$) along the $ab$-plane, \ie perpendicular to the helical axis, ferromagnetic domains are created between the winds of the helix, increasing the length of the magnetic unit cell and forming the chiral soliton lattice phase \cite{Togawa2012}. As field is increased, all spins become polarized  at $H_p^{ab}=0.18$ T.  Alternatively, when $H$ is applied along the $c$-axis, \ie along the direction of the helical wave vector, the helices smoothly evolve through a conical state.  The conical angle decreases with increasing $H$ until polarization at  $H_p^c =  2.5$ T~\cite{Miyadai1983,Ghimire2013}.   
Its electrical conduction is  quasi two-dimensional: the electrical resistivity measured with current flowing along the $c$-axis is  in the order of  $10^1-10^2$ times larger than with current in the $ab$-plane \cite{Ghimire2013, Togawa2013}.  In these regards, \crnbs$ $ displays similar magnetic and structural anisotropies as fabricated planar devices. 

So far,  SOC in magnetic multilayers or superlattice has been  studied intensively  in terms of engineering the electronic structure in order to control the spontaneous magnetic anisotropy \cite{Wang1993,Gimbert2012,Hotta2013}. Here, we demonstrate the reverse process in a similarly anisotropic layered  system \crnbs,  such that the  spin polarization direction controlled by applied magnetic field  alters the electronic structure via SOC.  For spin polarization along the $c$-axis, this results in enhanced  electrical conductivity, by increasing the DOS near the Fermi surface, as supported by our first principles calculations. This observation is the most prominent for temperatures $T<T_A$; $T_A$ is the temperature scale below which spin-disorder scattering contributes negligibly to the resistance, and is 
 empirically determined from the temperature dependence of the magnetoresistance.
Surprisingly,  roughly the  same  temperature scale of $T_A$ was found in the transverse  Hall resistivity, and is also consistent with the onset of deviation from Bloch's $T^{3/2}$ law for the magnetization \cite{Miyadai1983}. While $T_A$ is a crossover temperature and is  not sharply defined, it corresponds to the energy scale associated with modifications of the electronic structure, and is thus a useful quantity to estimate. 
%Our first  principle calculations illustrate the consistent result with the experimental observations. 

%\noindent{\bf {Experimental Methods}} 
Single crystals of \crnbs~were grown by vapor-transport (see Ref.~\cite{Ghimire2013} for details on growth and sample characterization).  Crystals from two different batches were used in this study, and show slightly different $T_C$s of 120 K and 133 K, but otherwise exhibit very little qualitative difference in the $T$ and $H$ dependences of both electrical and magnetic measurements.  All the data presented in this letter are from the sample with $T_C = 133$ K.  Samples from both batches were characterized with $x$-ray diffraction for the crystalline structure and small angle neutron scattering to verify the helimagnetic ground state~\cite{GhimireThesis}.   Magnetization was measured in a superconducting quantum interference device magnetometer.  Electrical transport properties were studied using standard four-probe measurements on samples with typical area $\approx 2\times 0.3$ mm$^2$ and thickness $\approx 20 - 70$ $\mu$m.  Electrical contacts, made by silver paint, had typical contact resistance less than 1 $\ohm$. Rotation of the sample in a field was performed by a home-built rotation probe, which was inserted in a superconducting split coil magnet.  For transport measurements, currents of $2-4$ mA were applied in the $ab$ plane. The magnetization measurements were performed using SQUID. We used  small wedges made of stycast epoxy with various angles  to mount the samples under angled magnetic field in magnetization measurements.
For the Hall measurement, we  checked the uniformity of the current flow by using multiple sets of electric contacts within a sample. The Hall data  is antisymmetrized  in order to remove  magneoresistance component  caused by a slight misalignment of the contacts.

The measurement is configured such that the flow of current lies within the crystallographic $ab$-plane, while the saturated magnetization sweeps the out-of-plane polar angle from $\theta_H = 0$ to $90^{\circ}$ (see the inset of Fig.\ref{angleMR}~(a)). 
This corresponds to the out-of-plane anisotropic magnetoresistance configuration as shown in \cite{Rijks1997, Kobs2011}. It is distinct from the typical AMR configuration where both  current and the magnetization lie within the film plane \cite{McGuire1975}. 
Note that the direction of current $always$ remains within the $ab$-plane and therefore, when $\theta_H=90 ^{\circ}$ the current is parallel to $H$.

%Figure1
\begin{figure}[htb]
\begin{center}
\includegraphics[width=0.85\linewidth]{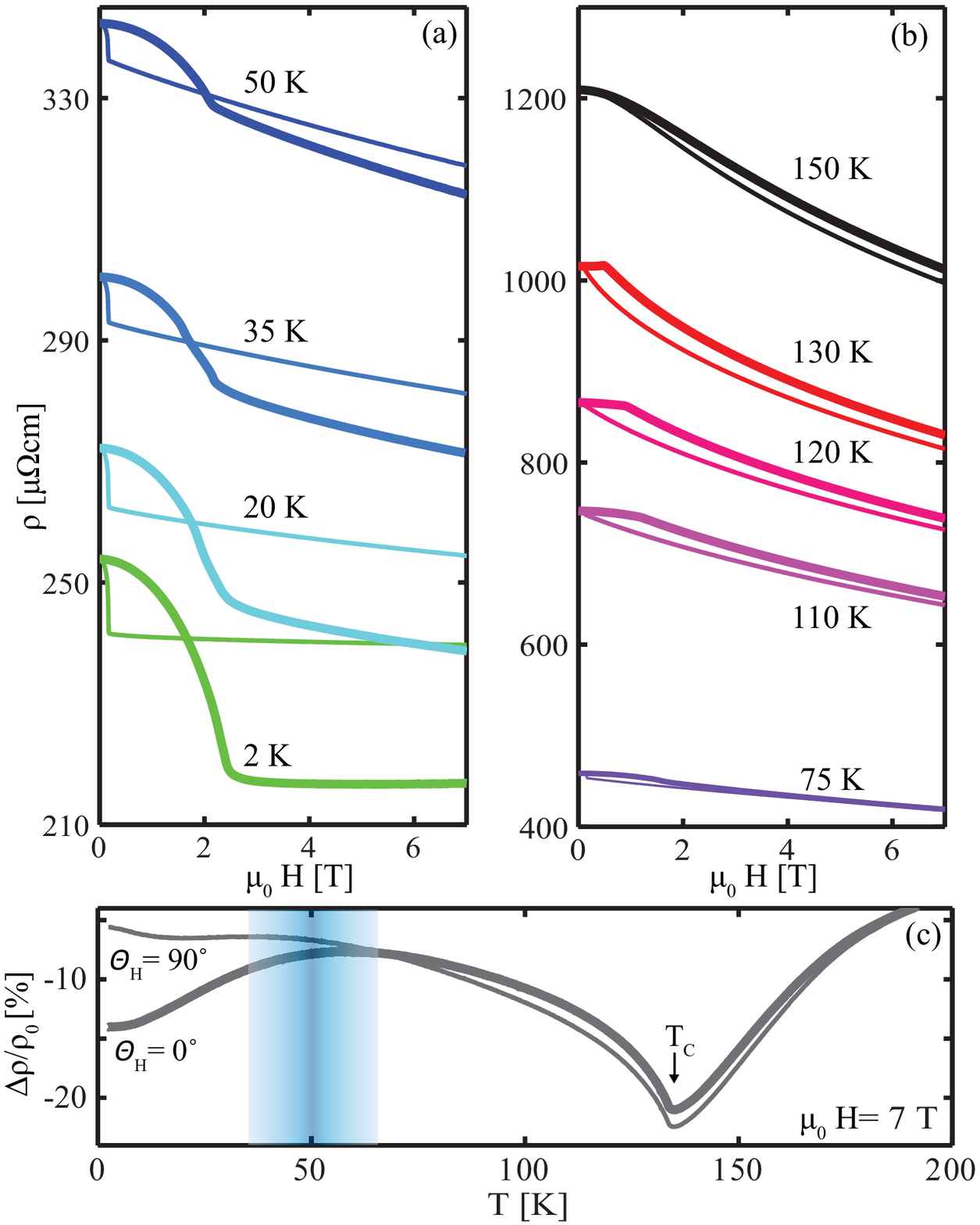}%{MRfMRM_fig1.eps}
\caption {\small  (a,b) In-plane resistivity $\rho$ as a function of $H$ in two different orientations: 
$\theta_H = 0^{\circ}$ (thick lines)  and $\theta_H = 90^{\circ}$(thin lines)  at fixed $T$s (a) $T \leq 50$ K  and (b) $T \geq 75$ K.  At each $T$, $H_p^c$ is clearly visible in both panels but in (b) $H_p^{ab}$ is hard to identity because of the large $y$-axis scale. (c) The normalized in-plane MR,  $\Delta\rho/\rho_0  = [\rho(H)-\rho_0]/\rho_0$ with $\rho_0= \rho(H=0)$ as a function of  $T$ measured at  $\mu_0H = 7$ T.   $T_C = 133$ K is marked with an arrow. Blurred blue line is centered at  $T_A = 52$ K, which is determined from the temperature at which $\rho(\theta_H=0,H_p^c)$ is no longer smaller that  $\rho(\theta_H=90{^\circ})$. However, since we intend $T_A$ to  indicate approximate $T$ scale above which the difference between the two MRs diminishes, we refer to  $T_A\approx 50$ K in the main text.
}
\label{MR}
\end{center}
\efig

First,  we plot the in-plane resistivities ($\rho$)  as a function of $H$  for  $\theta_H = 0$ (thick lines) and $\theta_H=90^{\circ}$ (thin lines)  at  different fixed temperatures ($T$)  in Fig.~\ref{MR}(a) and (b). 
Fig.\ref{MR}(c) summarizes the $T$ dependence of the in-plane magnetoresistance (MR), defined as  $\Delta\rho/\rho_0 =[\rho(H,\theta_H)-\rho_0]/\rho_0$ with $\rho_0= \rho(H=0)$, at $\mu_0  H=7$ T, where all of  the spins are  polarized along either orientation of $H$. The discrepancy of the in-plane  MR between the two orientations is most striking at 2 K,  where the reduction of MR  with out-of-plane magnetization is three times larger than with in-plane magnetization. We will come back to this below.  Also the $H$-independent resistivity values  in  $H> H_p^c$ and  $H > H_p^{ab}$  for both $H$ directions imply that all spins are polarized and the contribution of spin-disorder induced scattering is very little.

As $T$ increases, this trend reverses around $T_A \approx 50$ K, and $\rho(H, \theta_H=0^\circ)$ becomes  higher in  the entire $H$ range. 
$T_A$ is empirically determined from the temperature at which $\rho(\theta_H=0,H_p^c)$ is no longer smaller than  $\rho(\theta_H=90{^\circ}$), as shown in Fig. \ref{MR}(c).  In $T>T_A$,  the resistivity discrepancy between two spin polarization is overwhelmed by spin-disorder scattering induced  resistance.  This is also clear from  the $H$ dependence of two resistivities are parallel to each other in $T>T_A$ in Fig.~\ref{MR}(b).

The out-of-plane polarization field $H_p^c$  is clearly identified up to  $T _C$, as is $H_p^{ab}$, although the  large $x$-axis scale  makes it hard to identify in Fig.~\ref{MR}(b).  
At any given $T$,   the lower polarization field for $\theta_H =90^{\circ}$ indicates  a more effective suppression of spin-disorder for the same strength of $H$. This also gives rise to a  lower $\rho$ value in entire $H$ range  for high $T$s.
 The similar rate of  reduction of $\rho$   at higher $T$, which makes the two curves  at each temperature almost parallel, implies that  their $H$ dependence is attributed to the suppression of spin disorder.

%Figure2 
\begin{figure}[htb]
\begin{center}
\includegraphics[width=0.85\linewidth]{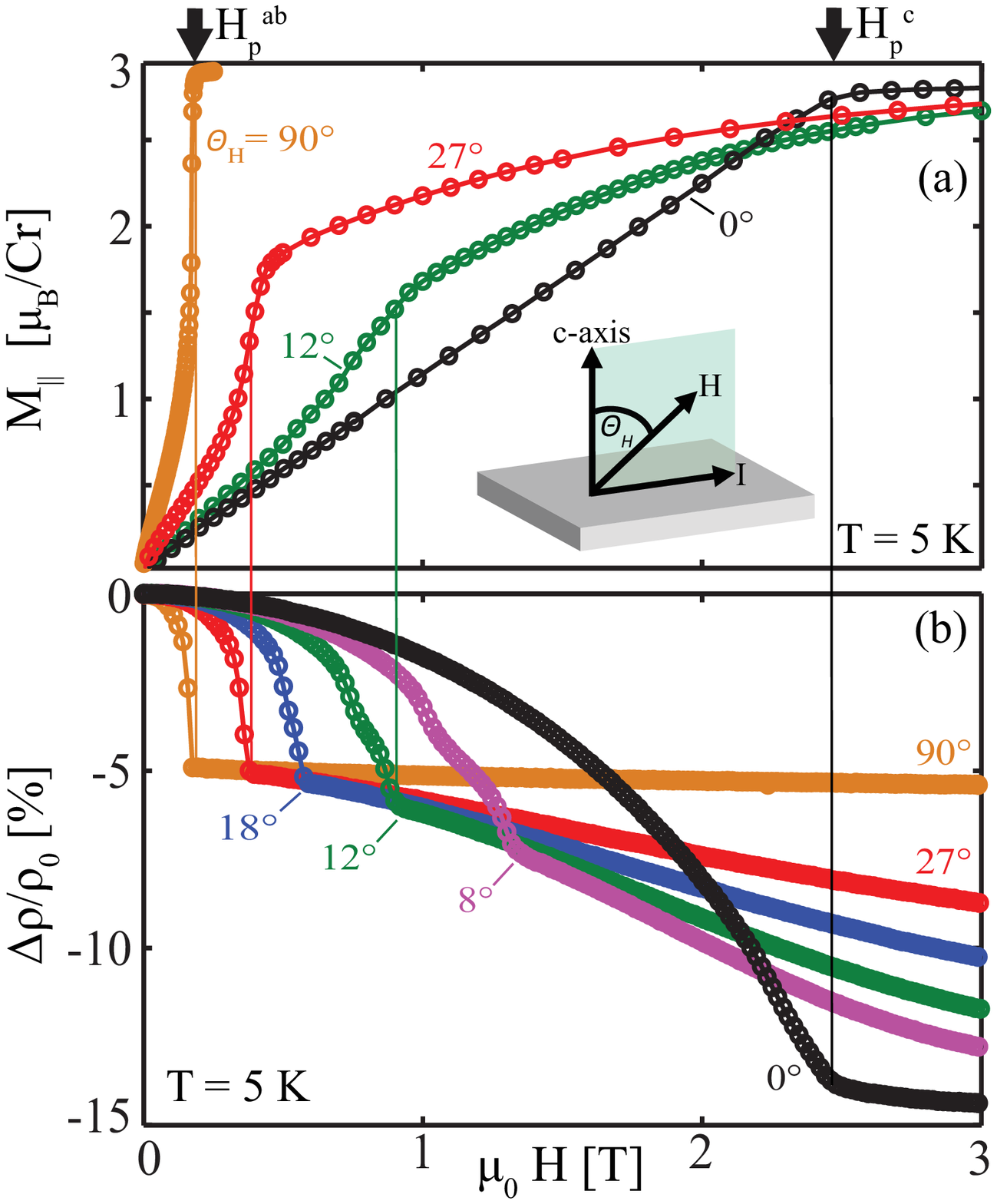}
\caption {\small 
(a) The magnetization component parallel to the applied magnetic field ($M_{\parallel}$) as a function of $H$ at different angle $\theta_H$s  at $T= 5$ K. The measurement configuration is shown in the inset.  
(b) In-plane MR as a function of $H$ oriented at angle $\theta_H$ and $T = 5$ K. Vertical lines mark the polarization field ($H_p^{\theta_H}$)
  at a given $\theta_H$. At intermediate angles, all spin are polarized at $H_p^{\theta_H}$ but  are not yet aligned along $H$ direction \cite{Chapman2014}. }
\label{angleMR}
\end{center}
\efig

In Fig.\ref{angleMR} (a) we plot the magnetization along the $H$ direction ($M_{\parallel}$) as a function of $H$ for different $\theta_H$s, at $T = 5$ K $\ll$ T$_C$.  The polarization fields ($H_p^{\theta_H}$)   are clearly visible, indicated with vertical lines and with arrows for both end angles of $\theta_H= 0$ and $90^{\circ}$. It occurs at fields well approximated by $H_p^{\theta_H} \simeq \frac{H_p^{ab}}{\sin\theta_H}$~\cite{Chapman2014}.
The rapid rise in magnetization observed  at large angles e.g. $\theta_H=90 ^{\circ}$ or $27^{\circ}$  indicates a  phase transition from the chiral soliton lattice to the ferromagnetic ordering.  At intermediate angles,  the rise at $H_p^{\theta_H}$ is not as rapid as at $\theta_H = 90^{\circ}$. While all spins are polarized  when $H=H_p^{\theta_H}$ at these angles, they are not yet aligned to the direction of $H$, but remain closer to the $ab$-plane. Upon increasing $H$ more, the polarized spins collectively rotate until they eventually align to $H$. This results in the gradual rise of  $M_{\parallel}$ in $H> H_p^{\theta_H}$ toward the saturated value, which is ascribed to the competition between Zeeman and the magnetic anisotropy energy \cite{Chapman2014}.

In Fig.~\ref{angleMR}~(b), we compare this with  the in-plane MR.
At $\theta_H = 90^{\circ}$, the reduction in MR by 5.6\% indicates a decrease in spin scattering as the helical ordering becomes polarized along $H$.  A clear kink  denotes the spin-polarization field $H_p^{ab}$. This dome-like shape is commonly observed  in other helimagnets for  the same  reason \cite{Lee2007, Chapman2013}.  
Once the spins completely polarize within the plane ($H > H_p^{ab}$), there is little variation in MR with $H$, which ensures that the contribution of the  spin-disorder induced MR is negligible. 
It is worth pointing out that $\Delta\rho/\rho_0$ at  $\theta_H= 90^{\circ}$  bears great similarity, in both  $H$ dependence and magnitude, to that of the interlayer MR, i.e. the MR when $I$ is applied along the $c$ axis and $H$ is in the $ab$ plane, reported in \cite{Togawa2013}. 

When $\theta_H$ approaches $0$, upon increasing $H$,  the MR decreases much slower initially but eventually surpasses the in-plane value, reaching down to 14.1\%, almost 3 times more reduction than $\theta_H =90 ^{\circ}$. 
This reduction is  remarkable  compared to  the variation found in a typical  traditional AMR phenomena, which are only a few tenths to a couple of percent at most \cite{McGuire1975}. 
 Although slightly larger than a couple of percent  out-of-plane AMR values have been reported in magnetic thin films systems \cite{Rijks1997}, it is attributed to the  reduction of spin scattering due to  the geometrical size and  texture of films. It is only very recent that the interface effect between adjacent magnetic and non-magnetic film was identified for a possible source of the  out-of-plane AMR effect, yet the reported size is less than  1\% \cite{Kobs2011}.

For $\theta_H= 8^{\circ}$ and $12^{\circ}$ there is another kink in the MR, appearing in the range of $H$ where the phase transition from chiral soliton lattice to ferromagnetic ordering occurs \cite{Togawa2013}.  This finite width of  the transition, only occurring for angles close to the $c$-axis, is believed to be caused by the formation of multiple domains undergoing  the transition at slightly different fields, as also seen in $M_{\parallel} (H)$  of Fig.\ref{angleMR}~(a).  While these kinks indicate a great sensitivity of the electrical transport to changes in the spin structure, it does not account for the continued decrease in MR in $H>H_p^{\theta_H}$.
%Fig3
\begin{figure}[htb]
\begin{center}
\includegraphics[width=0.85\linewidth]{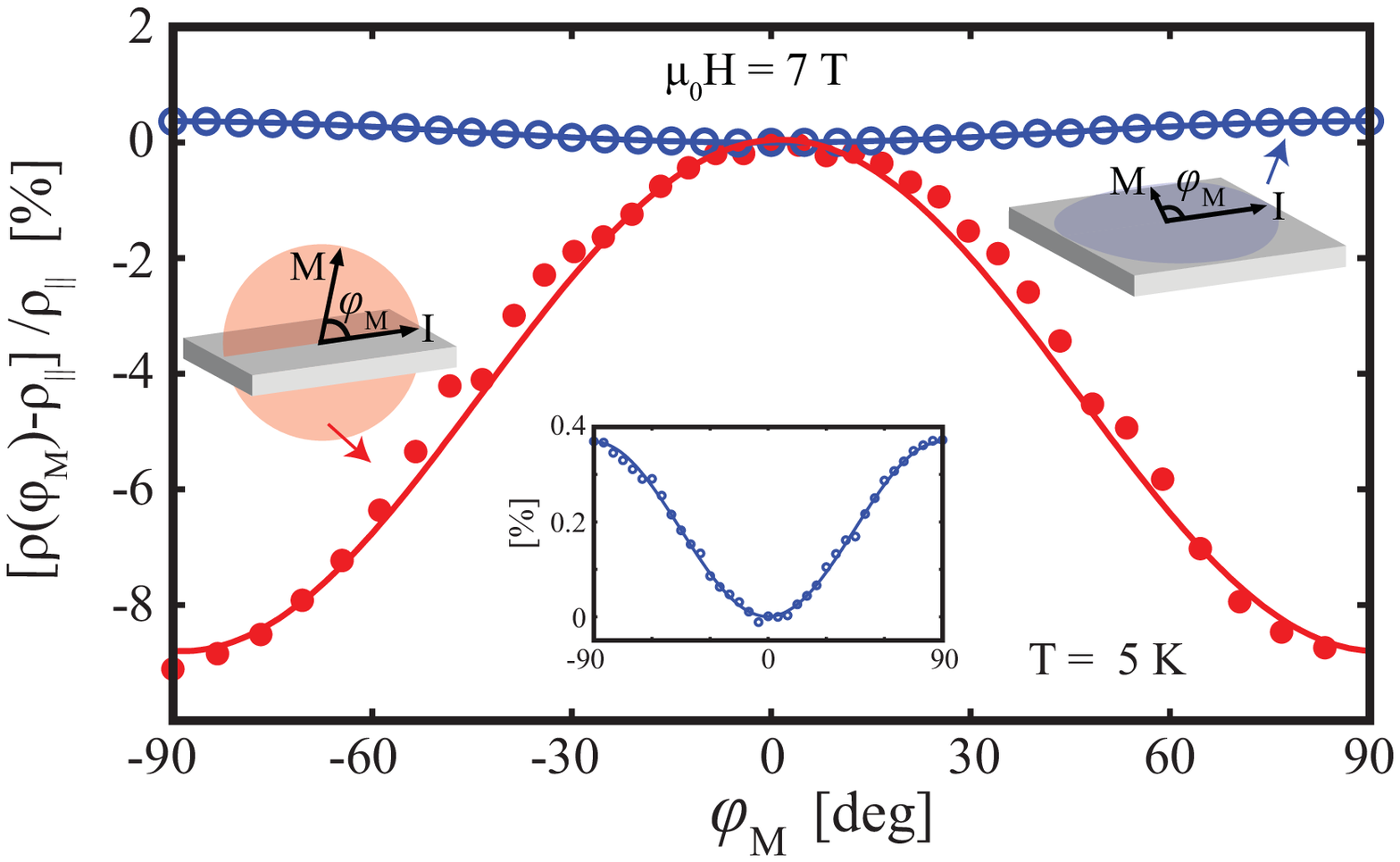}
\caption {\small 
 In-plane MR at $\mu_0H=7$ T, normalized with $\rho_{\parallel} (\varphi_M=0, \mu_0H=7~T)$,where  $\varphi_M$ denotes the angle between the current and $M$  both in-plane (open)  and out-of-plane (closed circle), as shown in the sketches. Note  the amplitude of out-of-plane AMR is 23 times larger  than in-plane AMR.
The angular dependences for both cases are found  $\propto \cos^2\varphi_M$, of which  is displayed in the solid lines. Inset is plotted  in-plane AMR with magnified $y$-axis.
}
\label{AMR}
\end{center}
\efig

To have a better understanding of this spin-orientation dependent  in-plane MR, we examine the AMR effect in both in-plane and out-of-plane rotation of applied field.  Schematics of the measurement are shown in insets of  Fig.\ref{AMR}.  In the main panel,  both in-plane and out-of-plane angular dependence of the in-plane resistivity normalized with $\rho_{\parallel} = \rho(\varphi_M=0, 	\mu_0H=7~T)$  are plotted, where $\varphi_M$ denotes the angle between $M$ and $I$ (see Supplementary Material).
In both cases, rotation of the magnetization  give rise to a $\cos^2\varphi_M$  angular dependences indicated with solid lines. However, the normalized  AMR  oscillation amplitude for the out-of-plane MR is larger by  23 times than the in-plane one;  +0.4\% and -9.1\%  change relative to $\rho_{\parallel}$ for in-plane and out-of-plane respectively. Note the different sign indicates the most reduced resistivity occurs when $M \parallel I$ for in-plane and $M\perp I$ for out-of-plane.
This out-of-plane AMR effect should also be distinguished from giant magnetoresistance of magnetic multilayers, in which  the  saturated MR remain the same for both orientations~\cite{Babibich1988}. 

%Fig4
\begin{figure}[htb]
\begin{center}
  \includegraphics[width=0.85\linewidth]{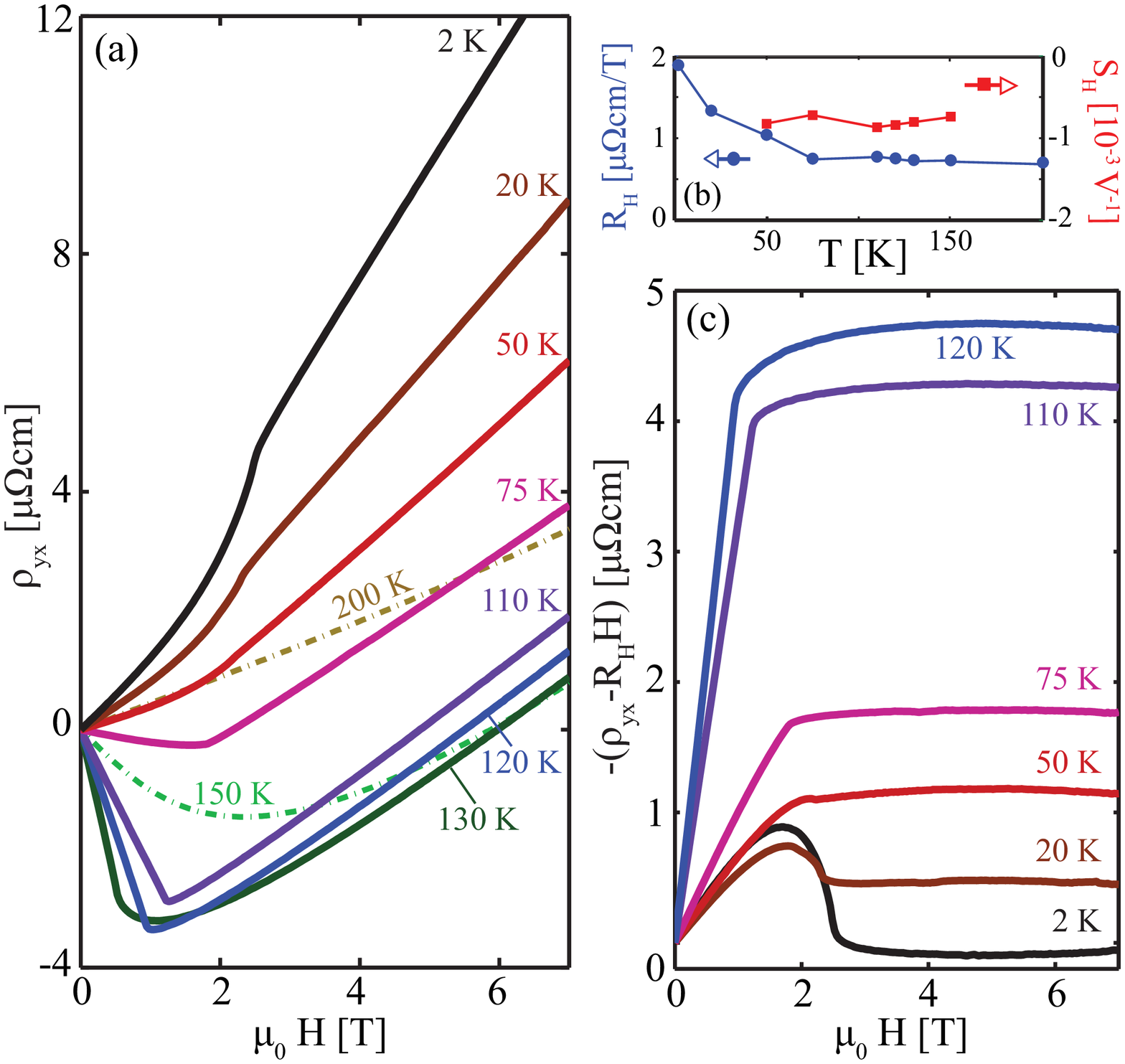}
  \caption {\small   (a)  $H$ dependence of $\rho_{yx}$  at various $T$s below  (solid lines) and  above (broken lines) $T_C$.  Peculiar concave  $H$-dependence occurs only in $T<T_A \approx 50$ K. Since $H \parallel c$,  all spins get $smoothly$ polarized along $c$-axis via helical and  conical state as $H$ increases. (b) The ordinary ($R_H$) and anomalous ($S_H$) Hall coefficients, determined from fits of $\rho_{yx}$.  Above $T_A$, both $R_H$ and $S_H$ are independent of $T$. Below $T_A$, the fit no longer works and $S_H$ cannot be obtained in a reliable way. $R_H$ is extracted from  $\mu_0H> 3$ T where $\rho_{yx}$ is linear to $H$. 
   (c) $H$ dependence of $\rho_{yx}$ less the OHE. AHE contribution decreases as $T$ is lowered and  is replaced with the hump-like feature in $T<T_A $, which has very little resemblance with $M(H)$.  The curve for 50 K displays a little dip before becoming flat, displaying the crossover between the hump-like shape and AHE component. 
  }
\label{Hall}
\end{center}
\efig

Finally, we examine the Hall effect which is measured  with  $H$ applied along $c$-axis ($\theta_H = 0$). This  configuration matches those in which a topological Hall effect (THE) was observed in other helimagnets, e.g. with $H$ along the helical axis ~\cite{Chapman2013}. Fig.~\ref{Hall}~(a) shows field sweeps of the Hall resistivity ($\rho_{yx}$) at various fixed $T$s.   
At low temperatures, an unexpected  $H$ dependence of $\rho_{yx}$ emerges, characterized by the pronounced concave curvature for $0<H<H_p^c$.   As shown in Fig.\ref{Hall}(c), this peculiar dependence is  more obvious after subtracting the ordinary Hall contribution, as described in the forthcoming analysis.  Note that this behavior emerges below $T_A \approx 50$ K, coinciding with  the emergence of the MR discrepancy described above.

For $T >T_A$, the Hall signal exhibits archetypical anomalous Hall behavior, empirically expressed as 
$\rho_{yx} (H) = \mu_0R_HH +\mu_0R_sM$, where $\mu_0$ is the vacuum permeability, $R_H$ normal Hall coefficient, and R$_s$ the anomalous Hall coefficient.  The first and second terms correspond to the ordinary Hall effect (OHE) and the anomalous Hall effect (AHE), respectively.  
As $T$ decreases, the AHE signal becomes visible at $T= 150$ K, and grows more prominent through $T_C$, where it is marked by a sharp kink at $H = H_p^c$. From the shape of the curves $R_S$ has the opposite sign as $R_H$.  
In the limit of the intrinsic AHE, we can rewrite the anomalous Hall coefficient as $R_S = S_H\rho^2(T,H)$, with $S_H$ an $H$ independent constant~\cite{Lee2007}, and fit our $\rho_{yx}(H)$ traces  well  in $T_A<T \le T_C$. 
The fitting parameters $R_H$ and $S_H$ are plotted in Fig.\ref{Hall}~(b). The effective carrier density derived from $R_H$ in $T>T_A$ is found  $9 \times 10^{20}$ holes/cm$^{3}$. It is consistent with the value found above $T_C$ at 200 K, where the Hall signal recovers almost a linear $H$ dependence. 

In $T<T_A$ however, this analysis is no longer valid and cannot capture  the concave $H$ dependence, even when an additional extrinsic contribution term is included. 
As seen in Fig.\ref{Hall}~(c), the emergence of a novel $H$ dependence below $H_p^c$  replaces the gradual reduction of the AHE  contribution.  It is hard to connect it with the AHE picture, as it bears no resemblance to $M (H)$.  
We  rule out the possibility that this behavior is caused by the topological Hall effect (THE) \cite{Lee2009,Neubauer2009} for the following reasons: 
first our spin structure model found  the skyrmion density is zero in this material for any temperature and the formation of skyrmions is unlikely due to the large magnetic anisotopy \cite{Chapman2014}. 
Second, the $T$ dependence -- occurring at low $T$s and vanishing as $T$ increases -- is the opposite to what was observed in the THE in other magnetic systems~\cite{Chapman2013}.
Lastly,  a recent neutron scattering study could not identify the existence of a complex periodic spin structure of skyrmions~ \cite{GhimireThesis}. 

It is interesting to compare the Hall effect of \crnbs~to that of Fe$_{1/4}$TaS$_2$, an anisotropic  magnetic dichalcogenide with a similar crystalline structure. 
 Both  have  the same  resistivity at 5 K, but \crnbs~has one-tenth the carrier density of Fe$_{1/4}$TaS$_2$. This means  the \crnbs~ has a ten-fold  larger mean free path ($l$).  The intrinsic AHE signal scales with $1/l^2$ and diminishes rapidly as $T$ is lowered and $\rho$ decreases \cite{Lee2007}, resulting in  the OHE dominating the Hall signal at  low temperatures, i.e. a recovery of $H$ linear dependence of $\rho_{yx}$.  This is consistent with our observation that the extrinsic AHE is negligible over the entire temperature range in \crnbs \cite{AHEnote}.

The fact that unusual $H$ profile of $\rho_{yx}$ and the larger amplitude of out-of-plane AMR occur in the same $T$ range, $T<T_A\approx 50$ K, implies that the two phenomena share a similar origin. In this temperature regime $R_H$, which was estimated from the slope of $\rho_{yx}$ when $H> 3$ T, increases rapidly.  The crossover temperature $T_A$ also appeared in other measurements: the $T$ dependence of the thermopower (see Supplementary Material) changes slowly just below $T_C$, and then after a broad shoulder around 50 K rapidly decreases.
 These unusual spin-orientation dependent transport features, unique to \crnbs, emerge only below $T_A$. $T_A$ is also consistent with a deviation from Bloch's $T^{3/2}$ law  of $M (T$)\cite{Miyadai1983}.  These observations points to a spin-orbit coupling effect of order 
$k_B T_A$, with   $k_B$  the Boltzmann constant.

%Fig5

\begin{figure}[htb]
\begin{center}
  \includegraphics[width=0.85\linewidth]{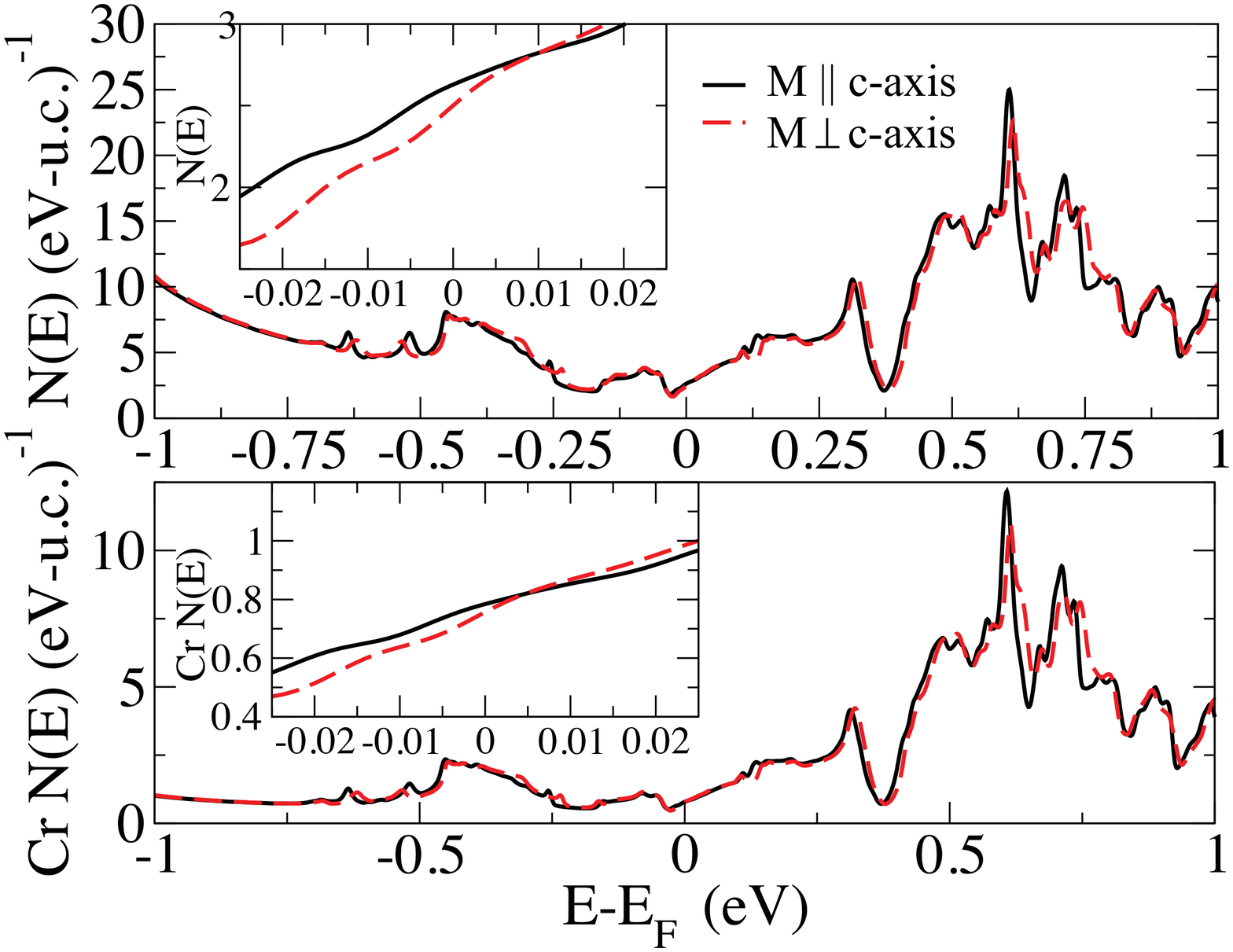}
  \caption {\small The calculated magnetic density-of-states of Cr$_{1/3}$NbS$_{2}$ for the indicated magnetic moment orientations; (a) total DOS, (b)  Cr DOS.  The energy scale $\pm$ 25 meV  centered at  Fermi energy $E_F$ is magnified in the insets.   The DOS at $E_F$ was found to be 3.2 \% greater when the moments orient along the c-axis instead of perpendicular to it, while the in-plane plasma frequencies are essentially unaffected by magnetization orientation, which is consistent with enhanced conductivity within an order of unity.  }
\label{fig:DOS}
\end{center}
\efig
\begin{figure}[htb]
\begin{center}
  \includegraphics[width=0.90\linewidth]{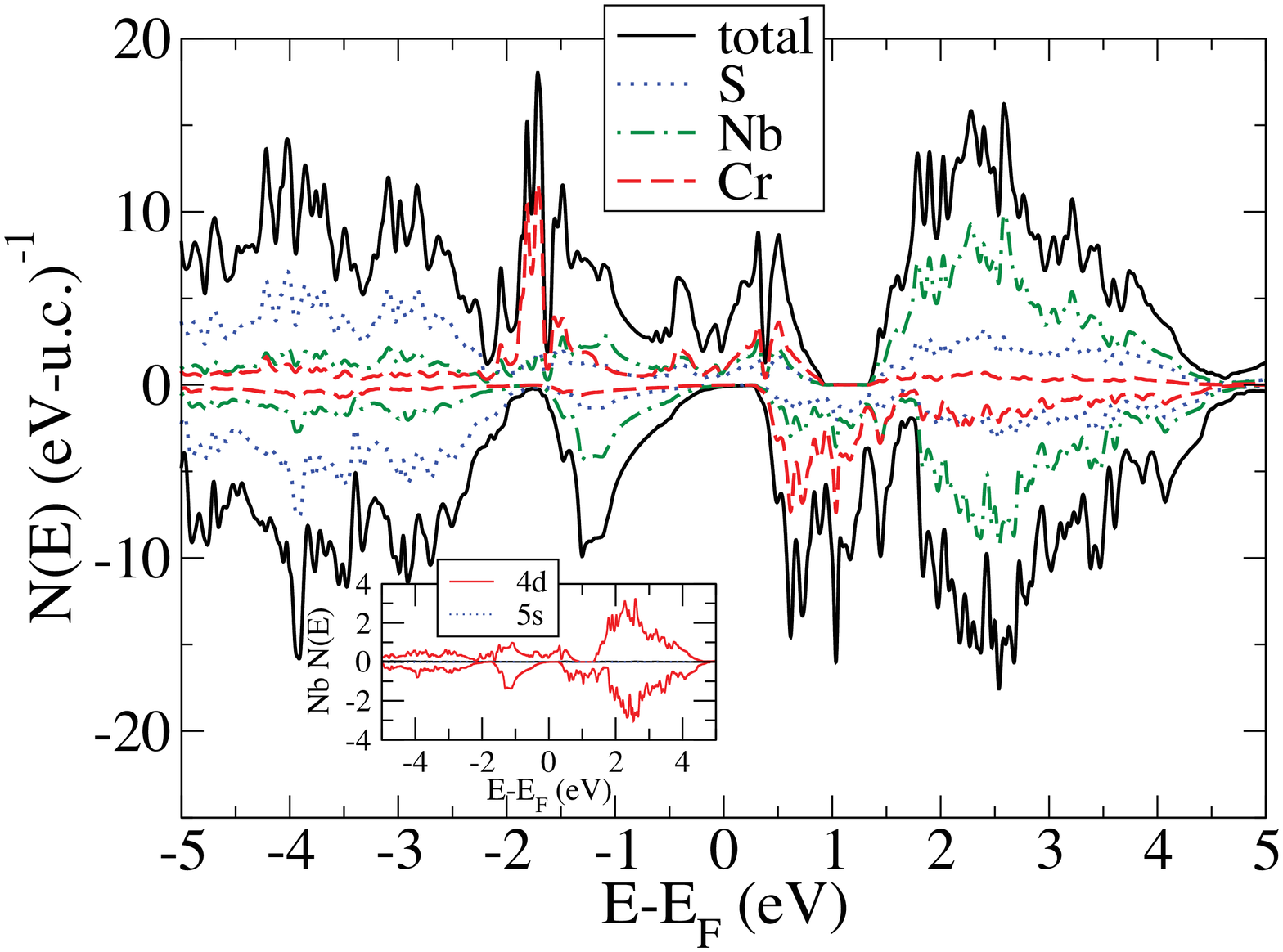}
  \caption {\small  The calculated magnetic density-of-states of Cr$_{1/3}$NbS$_{2}$ projected by spin and atomic character.  Inset: the orbital-projected character of the Nb DOS. }
\label{fig:DOS2}
\end{center}
\efig

In an attempt to understand these anisotropic magnetotransport properties, first principles calculations were pursued (procedural details in the Supplementary Material). Results of these calculations, depicted in Fig. 5  show a significant dependence of the calculated magnetic density of states on moment orientation, both for the total DOS (top) and Cr DOS (bottom). As indicated in the inset to Fig. 5, for states very near the Fermi level (E$_{F}$) -- less than 25 meV from E$_{F}$ - the DOS can vary by as much as 15\% depending on moment orientation.  Moreover, the Fermi level density of state (DOS) was found to be 3.2 \% greater when the moments orient along the $c$-axis instead of perpendicular to it, while  the in-plane plasma frequencies are essentially unaffected by magnetization orientation.  Thus, neglecting spin-orientation dependent scattering processes, the increase in DOS is largely responsible for the reduction of  the in-plane resistivity when the magnetization lies parallel to the $c$-axis instead of the $ab$-plane.  The resistivities for the two spin-polarization directions  at  5 K  differ by 11\%, which is bigger than  the DOS difference by a factor of the  order of unity. Also, by examining the bands near $E_F$ (within a 20 meV window) for both orientations, one finds the averaged offset in the energies to be approximately 1.7 meV $\simeq$ 20 K. This is a direct result of SOC and consistent with the temperature scale $T_A$ found in the MR  and Hall effect data. These changes in electronic structure and their subsequent modification of scattering processes are likely responsible for the spin orientation dependence in the transport. 

We note also from Fig. 5 that the Cr-site DOS is of essentially the same shape as the total DOS, within an eV of E$_{F}$.  In addition, the effects on the Cr DOS of the change in moment orientation from [001] ($\parallel$ $c$)  to [100] ($\parallel$ $ab$) closely parallel the changes in the total DOS.  This is due to the primary role of the Cr atoms in the magnetic behavior - only for atoms with a large local moment, such as the Cr here, does one expect a moment-orientation dependent DOS. Note that in the calculation the induced Nb moment is only 0.05 $\mu_B$ and that on S essentially zero. 

In Figure 6 we show the magnetic state DOS projected into spin-up and spin-down sub-bands, along with the Nb $4d$ and $5s$ orbital-projected DOS.  A substantial exchange splitting is visible in the Cr DOS, as expected given its magnetic character, with some minor splitting also observed in the Nb and S.  There is little, if any, spin-down character to the Cr DOS for nearly 2 eV below E$_{F}$.  This is suggestive of the strength of the magnetism in this system, as indicated by the 3 $\mu_B$/Cr total magnetic moment found both in the calculation and the experiment.  There is substantial hybridization around $E_{F}$, indicative of covalent bonding, although the Cr states are generally confined to within 2 eV of E$_{F}$ and the Nb and S predominate outside this range.  In the inset of Figure 6 we plot the orbital-projected Nb DOS, which is dominated by $4d$ contributions, with the $5s$ contribution negligible.

Further detailed investigation on the electronic structure depending on the spin direction will shed light on  understanding  the difference of amplitudes of  in-plane and out-of-plane AMR. This mechanism is suspected to be closely related to  abnormal $H$ dependence of the Hall effect observed for $T<T_A$.

The remarkable sensitivity of the magnetotransport behavior to polarized spin orientation in \crnbs~  is mostly driven by Cr$^{3+}$ ions.  
Strong  magnetization orientation dependence of  the electronic structure of $3d$ metals leads to  large effects on  the magnetotransport properties in \crnbs. 
The same strong dependence also underlies efforts to control magnetic anisotropy in multilayer systems made of $3d$ magnetic elements.
%This is also connected  to the fact that  the majority of 
%This is not entirely unexpected as 3$d$ metals have exhibited a strong band structure dependence on the orientation of magnetization. 
%In fact, engineering  spin-orbit coupling  at the interface of magnetic multilayers is one of the primary routes to obtain magnetoanisotropy perpendicular to the film plane.
Our results  demonstrates a closely interwoven relation among the orientation of magnetization,  the  resultant electronic structure change and its influence on the electrical conduction, all of which are mediated by spin-orbit coupling. 
Furthermore,  the non-collinear  helical ground state enriches the role of spin-orbit coupling and the subsequent complexity of  the electrical transport properties.

\vspace{0.1in}

\vspace{0.1in}
\noindent{\bf {Acknowledgment}} The authors thank J. G. Checkelsky, X. Fan, M. Hermele and K. M. McElroy for enlightening discussions. 
This work was supported by the US DOE, Basic Energy Sciences, Materials Sciences and Engineering Division (ORNL) and at CU under Award Number DE-SC0006888.  D.P. is supported by the Critical Materials Institute, an Energy Innovation Hub funded by the U.S. DOE, Energy
Efficiency and Renewable Energy, Advanced Manufacturing Office.

\vspace{0.1in}
\noindent $^{\dagger}$Current Address: Los Alamos National Laboratory, Los Alamos, NM 
%

%\bibliography{chalcobib}

\begin{thebibliography}{30}
\expandafter\ifx\csname natexlab\endcsname\relax\def\natexlab#1{#1}\fi
\expandafter\ifx\csname bibnamefont\endcsname\relax
  \def\bibnamefont#1{#1}\fi
\expandafter\ifx\csname bibfnamefont\endcsname\relax
  \def\bibfnamefont#1{#1}\fi
\expandafter\ifx\csname citenamefont\endcsname\relax
  \def\citenamefont#1{#1}\fi
\expandafter\ifx\csname url\endcsname\relax
  \def\url#1{\texttt{#1}}\fi
\expandafter\ifx\csname urlprefix\endcsname\relax\def\urlprefix{URL }\fi
\providecommand{\bibinfo}[2]{#2}
\providecommand{\eprint}[2][]{\url{#2}}

\bibitem[{\citenamefont{Brooks}(1940)}]{Brooks1940}
\bibinfo{author}{\bibfnamefont{H.}~\bibnamefont{Brooks}},
  \bibinfo{journal}{Phys. Rev.} \textbf{\bibinfo{volume}{58}},
  \bibinfo{pages}{909} (\bibinfo{year}{1940}),
  \urlprefix\url{http://link.aps.org/doi/10.1103/PhysRev.58.909}.

\bibitem[{\citenamefont{Bruno}(1989)}]{Bruno1989}
\bibinfo{author}{\bibfnamefont{P.}~\bibnamefont{Bruno}},
  \bibinfo{journal}{Phys. Rev. B} \textbf{\bibinfo{volume}{39}},
  \bibinfo{pages}{865} (\bibinfo{year}{1989}),
  \urlprefix\url{http://link.aps.org/doi/10.1103/PhysRevB.39.865}.

\bibitem[{\citenamefont{Wang et~al.}(1993)\citenamefont{Wang, Wu, and
  Freeman}}]{Wang1993}
\bibinfo{author}{\bibfnamefont{D.-s.} \bibnamefont{Wang}},
  \bibinfo{author}{\bibfnamefont{R.}~\bibnamefont{Wu}}, \bibnamefont{and}
  \bibinfo{author}{\bibfnamefont{A.~J.} \bibnamefont{Freeman}},
  \bibinfo{journal}{Phys. Rev. B} \textbf{\bibinfo{volume}{47}},
  \bibinfo{pages}{14932} (\bibinfo{year}{1993}),
  \urlprefix\url{http://link.aps.org/doi/10.1103/PhysRevB.47.14932}.

\bibitem[{\citenamefont{Bode et~al.}(2002)\citenamefont{Bode, Heinze, Kubetzka,
  Pietzsch, Nie, Bihlmayer, Bl\"ugel, and Wiesendanger}}]{Bode2002}
\bibinfo{author}{\bibfnamefont{M.}~\bibnamefont{Bode}},
  \bibinfo{author}{\bibfnamefont{S.}~\bibnamefont{Heinze}},
  \bibinfo{author}{\bibfnamefont{A.}~\bibnamefont{Kubetzka}},
  \bibinfo{author}{\bibfnamefont{O.}~\bibnamefont{Pietzsch}},
  \bibinfo{author}{\bibfnamefont{X.}~\bibnamefont{Nie}},
  \bibinfo{author}{\bibfnamefont{G.}~\bibnamefont{Bihlmayer}},
  \bibinfo{author}{\bibfnamefont{S.}~\bibnamefont{Bl\"ugel}}, \bibnamefont{and}
  \bibinfo{author}{\bibfnamefont{R.}~\bibnamefont{Wiesendanger}},
  \bibinfo{journal}{Phys. Rev. Lett.} \textbf{\bibinfo{volume}{89}},
  \bibinfo{pages}{237205} (\bibinfo{year}{2002}),
  \urlprefix\url{http://link.aps.org/doi/10.1103/PhysRevLett.89.237205}.

\bibitem[{\citenamefont{Gimbert and Calmels}(2012)}]{Gimbert2012}
\bibinfo{author}{\bibfnamefont{F.}~\bibnamefont{Gimbert}} \bibnamefont{and}
  \bibinfo{author}{\bibfnamefont{L.}~\bibnamefont{Calmels}},
  \bibinfo{journal}{Phys. Rev. B} \textbf{\bibinfo{volume}{86}},
  \bibinfo{pages}{184407} (\bibinfo{year}{2012}),
  \urlprefix\url{http://link.aps.org/doi/10.1103/PhysRevB.86.184407}.

\bibitem[{\citenamefont{Hotta et~al.}(2013)\citenamefont{Hotta, Nakamura,
  Akiyama, Ito, Oguchi, and Freeman}}]{Hotta2013}
\bibinfo{author}{\bibfnamefont{K.}~\bibnamefont{Hotta}},
  \bibinfo{author}{\bibfnamefont{K.}~\bibnamefont{Nakamura}},
  \bibinfo{author}{\bibfnamefont{T.}~\bibnamefont{Akiyama}},
  \bibinfo{author}{\bibfnamefont{T.}~\bibnamefont{Ito}},
  \bibinfo{author}{\bibfnamefont{T.}~\bibnamefont{Oguchi}}, \bibnamefont{and}
  \bibinfo{author}{\bibfnamefont{A.~J.} \bibnamefont{Freeman}},
  \bibinfo{journal}{Phys. Rev. Lett.} \textbf{\bibinfo{volume}{110}},
  \bibinfo{pages}{267206} (\bibinfo{year}{2013}),
  \urlprefix\url{http://link.aps.org/doi/10.1103/PhysRevLett.110.267206}.

\bibitem[{\citenamefont{Muhlbauer et~al.}(2009)\citenamefont{Muhlbauer, Binz,
  Jonietz, Pfleiderer, Rosch, Neubauer, Georgii, and Boni}}]{Muhlbauer2009}
\bibinfo{author}{\bibfnamefont{S.}~\bibnamefont{Muhlbauer}},
  \bibinfo{author}{\bibfnamefont{B.}~\bibnamefont{Binz}},
  \bibinfo{author}{\bibfnamefont{F.}~\bibnamefont{Jonietz}},
  \bibinfo{author}{\bibfnamefont{C.}~\bibnamefont{Pfleiderer}},
  \bibinfo{author}{\bibfnamefont{A.}~\bibnamefont{Rosch}},
  \bibinfo{author}{\bibfnamefont{A.}~\bibnamefont{Neubauer}},
  \bibinfo{author}{\bibfnamefont{R.}~\bibnamefont{Georgii}}, \bibnamefont{and}
  \bibinfo{author}{\bibfnamefont{P.}~\bibnamefont{Boni}},
  \bibinfo{journal}{Science} \textbf{\bibinfo{volume}{323}},
  \bibinfo{pages}{915} (\bibinfo{year}{2009}).

\bibitem[{\citenamefont{Lee et~al.}(2009)\citenamefont{Lee, Kang, Onose,
  Tokura, and Ong}}]{Lee2009}
\bibinfo{author}{\bibfnamefont{M.}~\bibnamefont{Lee}},
  \bibinfo{author}{\bibfnamefont{W.}~\bibnamefont{Kang}},
  \bibinfo{author}{\bibfnamefont{Y.}~\bibnamefont{Onose}},
  \bibinfo{author}{\bibfnamefont{Y.}~\bibnamefont{Tokura}}, \bibnamefont{and}
  \bibinfo{author}{\bibfnamefont{N.~P.} \bibnamefont{Ong}},
  \bibinfo{journal}{Phys. Rev. Lett.} \textbf{\bibinfo{volume}{102}},
  \bibinfo{pages}{186601} (\bibinfo{year}{2009}).

\bibitem[{\citenamefont{Neubauer et~al.}(2009)\citenamefont{Neubauer,
  Pfleiderer, Binz, Rosch, Ritz, Niklowitz, and B\"oni}}]{Neubauer2009}
\bibinfo{author}{\bibfnamefont{A.}~\bibnamefont{Neubauer}},
  \bibinfo{author}{\bibfnamefont{C.}~\bibnamefont{Pfleiderer}},
  \bibinfo{author}{\bibfnamefont{B.}~\bibnamefont{Binz}},
  \bibinfo{author}{\bibfnamefont{A.}~\bibnamefont{Rosch}},
  \bibinfo{author}{\bibfnamefont{R.}~\bibnamefont{Ritz}},
  \bibinfo{author}{\bibfnamefont{P.~G.} \bibnamefont{Niklowitz}},
  \bibnamefont{and} \bibinfo{author}{\bibfnamefont{P.}~\bibnamefont{B\"oni}},
  \bibinfo{journal}{Phys. Rev. Lett.} \textbf{\bibinfo{volume}{102}},
  \bibinfo{pages}{186602} (\bibinfo{year}{2009}),
  \urlprefix\url{http://link.aps.org/doi/10.1103/PhysRevLett.102.186602}.

\bibitem[{\citenamefont{Dzyaloshinsky}(1958)}]{Dzyaloshinsky1958}
\bibinfo{author}{\bibfnamefont{I.}~\bibnamefont{Dzyaloshinsky}},
  \bibinfo{journal}{Journal of Physics and Chemistry of Solids}
  \textbf{\bibinfo{volume}{4}}, \bibinfo{pages}{241 } (\bibinfo{year}{1958}),
  ISSN \bibinfo{issn}{0022-3697},
  \urlprefix\url{http://www.sciencedirect.com/science/article/pii/0022369758900763}.

\bibitem[{\citenamefont{Moriya}(1960)}]{Moriya1960}
\bibinfo{author}{\bibfnamefont{T.}~\bibnamefont{Moriya}},
  \bibinfo{journal}{Phys. Rev.} \textbf{\bibinfo{volume}{120}},
  \bibinfo{pages}{91} (\bibinfo{year}{1960}),
  \urlprefix\url{http://link.aps.org/doi/10.1103/PhysRev.120.91}.

\bibitem[{\citenamefont{Hopkinson and Kee}(2009)}]{Hopkinson2009}
\bibinfo{author}{\bibfnamefont{J.~M.} \bibnamefont{Hopkinson}}
  \bibnamefont{and} \bibinfo{author}{\bibfnamefont{H.-Y.} \bibnamefont{Kee}},
  \bibinfo{journal}{Phys. Rev. B} \textbf{\bibinfo{volume}{79}},
  \bibinfo{pages}{014421} (\bibinfo{year}{2009}),
  \urlprefix\url{http://link.aps.org/doi/10.1103/PhysRevB.79.014421}.

\bibitem[{\citenamefont{Fert et~al.}(2013)\citenamefont{Fert, Cros, and
  Sampaio}}]{Fert2013}
\bibinfo{author}{\bibfnamefont{A.}~\bibnamefont{Fert}},
  \bibinfo{author}{\bibfnamefont{V.}~\bibnamefont{Cros}}, \bibnamefont{and}
  \bibinfo{author}{\bibfnamefont{J.}~\bibnamefont{Sampaio}},
  \bibinfo{journal}{Nature nanotechnology} \textbf{\bibinfo{volume}{8}},
  \bibinfo{pages}{152} (\bibinfo{year}{2013}).

\bibitem[{\citenamefont{Romming et~al.}(2013)\citenamefont{Romming, Hanneken,
  Menzel, Bickel, Wolter, von Bergmann, Kubetzka, and
  Wiesendanger}}]{Romming2013}
\bibinfo{author}{\bibfnamefont{N.}~\bibnamefont{Romming}},
  \bibinfo{author}{\bibfnamefont{C.}~\bibnamefont{Hanneken}},
  \bibinfo{author}{\bibfnamefont{M.}~\bibnamefont{Menzel}},
  \bibinfo{author}{\bibfnamefont{J.~E.} \bibnamefont{Bickel}},
  \bibinfo{author}{\bibfnamefont{B.}~\bibnamefont{Wolter}},
  \bibinfo{author}{\bibfnamefont{K.}~\bibnamefont{von Bergmann}},
  \bibinfo{author}{\bibfnamefont{A.}~\bibnamefont{Kubetzka}}, \bibnamefont{and}
  \bibinfo{author}{\bibfnamefont{R.}~\bibnamefont{Wiesendanger}},
  \bibinfo{journal}{Science} \textbf{\bibinfo{volume}{341}},
  \bibinfo{pages}{636} (\bibinfo{year}{2013}).

\bibitem[{\citenamefont{Sales et~al.}(2008)\citenamefont{Sales, Jin, and
  Mandrus}}]{Sales2008}
\bibinfo{author}{\bibfnamefont{B.~C.} \bibnamefont{Sales}},
  \bibinfo{author}{\bibfnamefont{R.}~\bibnamefont{Jin}}, \bibnamefont{and}
  \bibinfo{author}{\bibfnamefont{D.}~\bibnamefont{Mandrus}},
  \bibinfo{journal}{Phys. Rev. B} \textbf{\bibinfo{volume}{77}},
  \bibinfo{pages}{024409} (\bibinfo{year}{2008}),
  \urlprefix\url{http://link.aps.org/doi/10.1103/PhysRevB.77.024409}.

\bibitem[{\citenamefont{Mokrousov et~al.}(2013)\citenamefont{Mokrousov, Zhang,
  Freimuth, Zimmermann, Long, Weischenberg, Mavropoulos, and
  Blugel}}]{Mokrousov2013}
\bibinfo{author}{\bibfnamefont{Y.}~\bibnamefont{Mokrousov}},
  \bibinfo{author}{\bibfnamefont{H.}~\bibnamefont{Zhang}},
  \bibinfo{author}{\bibfnamefont{F.}~\bibnamefont{Freimuth}},
  \bibinfo{author}{\bibfnamefont{B.}~\bibnamefont{Zimmermann}},
  \bibinfo{author}{\bibfnamefont{N.~H.} \bibnamefont{Long}},
  \bibinfo{author}{\bibfnamefont{I.}~\bibnamefont{Weischenberg},
  \bibfnamefont{J.and~Souza}},
  \bibinfo{author}{\bibfnamefont{P.}~\bibnamefont{Mavropoulos}},
  \bibnamefont{and} \bibinfo{author}{\bibfnamefont{S.}~\bibnamefont{Blugel}},
  \bibinfo{journal}{J. Phys.: Cond. Matt.} \textbf{\bibinfo{volume}{25}},
  \bibinfo{pages}{163201} (\bibinfo{year}{2013}).

\bibitem[{\citenamefont{Ghimire et~al.}(2013)\citenamefont{Ghimire, McGuire,
  Parker, Sipos, Tang, Yan, Sales, and Mandrus}}]{Ghimire2013}
\bibinfo{author}{\bibfnamefont{N.~J.} \bibnamefont{Ghimire}},
  \bibinfo{author}{\bibfnamefont{M.~A.} \bibnamefont{McGuire}},
  \bibinfo{author}{\bibfnamefont{D.~S.} \bibnamefont{Parker}},
  \bibinfo{author}{\bibfnamefont{B.}~\bibnamefont{Sipos}},
  \bibinfo{author}{\bibfnamefont{S.}~\bibnamefont{Tang}},
  \bibinfo{author}{\bibfnamefont{J.-Q.} \bibnamefont{Yan}},
  \bibinfo{author}{\bibfnamefont{B.~C.} \bibnamefont{Sales}}, \bibnamefont{and}
  \bibinfo{author}{\bibfnamefont{D.}~\bibnamefont{Mandrus}},
  \bibinfo{journal}{Phys. Rev. B} \textbf{\bibinfo{volume}{87}},
  \bibinfo{pages}{104403} (\bibinfo{year}{2013}),
  \urlprefix\url{http://link.aps.org/doi/10.1103/PhysRevB.87.104403}.

\bibitem[{\citenamefont{Moriya and Miyadai}(1982)}]{Moriya1982}
\bibinfo{author}{\bibfnamefont{T.}~\bibnamefont{Moriya}} \bibnamefont{and}
  \bibinfo{author}{\bibfnamefont{T.}~\bibnamefont{Miyadai}},
  \bibinfo{journal}{Solid State Comm.} \textbf{\bibinfo{volume}{42}},
  \bibinfo{pages}{209} (\bibinfo{year}{1982}).

\bibitem[{\citenamefont{Miyadai et~al.}(1983)\citenamefont{Miyadai, Kikuchi,
  .Kondo, Sakka, Arai, and Ishikawa}}]{Miyadai1983}
\bibinfo{author}{\bibfnamefont{T.}~\bibnamefont{Miyadai}},
  \bibinfo{author}{\bibfnamefont{K.}~\bibnamefont{Kikuchi}},
  \bibinfo{author}{\bibfnamefont{H.}~\bibnamefont{.Kondo}},
  \bibinfo{author}{\bibfnamefont{S.}~\bibnamefont{Sakka}},
  \bibinfo{author}{\bibfnamefont{M.}~\bibnamefont{Arai}}, \bibnamefont{and}
  \bibinfo{author}{\bibfnamefont{Y.}~\bibnamefont{Ishikawa}},
  \bibinfo{journal}{J. Phys. Soc. Jpn.} \textbf{\bibinfo{volume}{52}},
  \bibinfo{pages}{1394} (\bibinfo{year}{1983}).

\bibitem[{\citenamefont{Togawa et~al.}(2012)\citenamefont{Togawa, Koyama,
  Takayanagi, Mori, Kousaka, Akimitsu, Nishihara, Inoue, Ovchinnikov, and
  Kishine}}]{Togawa2012}
\bibinfo{author}{\bibfnamefont{Y.}~\bibnamefont{Togawa}},
  \bibinfo{author}{\bibfnamefont{T.}~\bibnamefont{Koyama}},
  \bibinfo{author}{\bibfnamefont{K.}~\bibnamefont{Takayanagi}},
  \bibinfo{author}{\bibfnamefont{S.}~\bibnamefont{Mori}},
  \bibinfo{author}{\bibfnamefont{Y.}~\bibnamefont{Kousaka}},
  \bibinfo{author}{\bibfnamefont{J.}~\bibnamefont{Akimitsu}},
  \bibinfo{author}{\bibfnamefont{S.}~\bibnamefont{Nishihara}},
  \bibinfo{author}{\bibfnamefont{K.}~\bibnamefont{Inoue}},
  \bibinfo{author}{\bibfnamefont{A.~S.} \bibnamefont{Ovchinnikov}},
  \bibnamefont{and} \bibinfo{author}{\bibfnamefont{J.}~\bibnamefont{Kishine}},
  \bibinfo{journal}{Phys. Rev. Lett.} \textbf{\bibinfo{volume}{108}},
  \bibinfo{pages}{107202} (\bibinfo{year}{2012}),
  \urlprefix\url{http://link.aps.org/doi/10.1103/PhysRevLett.108.107202}.

\bibitem[{\citenamefont{Togawa et~al.}(2013)\citenamefont{Togawa, Kousaka,
  Nishihara, Inoue, Akimitsu, Ovchinnikov, and Kishine}}]{Togawa2013}
\bibinfo{author}{\bibfnamefont{Y.}~\bibnamefont{Togawa}},
  \bibinfo{author}{\bibfnamefont{Y.}~\bibnamefont{Kousaka}},
  \bibinfo{author}{\bibfnamefont{S.}~\bibnamefont{Nishihara}},
  \bibinfo{author}{\bibfnamefont{K.}~\bibnamefont{Inoue}},
  \bibinfo{author}{\bibfnamefont{J.}~\bibnamefont{Akimitsu}},
  \bibinfo{author}{\bibfnamefont{A.~S.} \bibnamefont{Ovchinnikov}},
  \bibnamefont{and} \bibinfo{author}{\bibfnamefont{J.}~\bibnamefont{Kishine}},
  \bibinfo{journal}{Phys. Rev. Lett.} \textbf{\bibinfo{volume}{111}},
  \bibinfo{pages}{197204} (\bibinfo{year}{2013}),
  \urlprefix\url{http://link.aps.org/doi/10.1103/PhysRevLett.111.197204}.

\bibitem[{\citenamefont{Rijks et~al.}(1997)\citenamefont{Rijks, Lenczowski,
  Coehoorn, and de~Jonge}}]{Rijks1997}
\bibinfo{author}{\bibfnamefont{T.~G. S.~M.} \bibnamefont{Rijks}},
  \bibinfo{author}{\bibfnamefont{S.~K.~J.} \bibnamefont{Lenczowski}},
  \bibinfo{author}{\bibfnamefont{R.}~\bibnamefont{Coehoorn}}, \bibnamefont{and}
  \bibinfo{author}{\bibfnamefont{W.~J.~M.} \bibnamefont{de~Jonge}},
  \bibinfo{journal}{Phys. Rev. B} \textbf{\bibinfo{volume}{56}},
  \bibinfo{pages}{362} (\bibinfo{year}{1997}),
  \urlprefix\url{http://link.aps.org/doi/10.1103/PhysRevB.56.362}.

\bibitem[{\citenamefont{Kobs et~al.}(2011)\citenamefont{Kobs, He\ss{}e,
  Kreuzpaintner, Winkler, Lott, Weinberger, Schreyer, and Oepen}}]{Kobs2011}
\bibinfo{author}{\bibfnamefont{A.}~\bibnamefont{Kobs}},
  \bibinfo{author}{\bibfnamefont{S.}~\bibnamefont{He\ss{}e}},
  \bibinfo{author}{\bibfnamefont{W.}~\bibnamefont{Kreuzpaintner}},
  \bibinfo{author}{\bibfnamefont{G.}~\bibnamefont{Winkler}},
  \bibinfo{author}{\bibfnamefont{D.}~\bibnamefont{Lott}},
  \bibinfo{author}{\bibfnamefont{P.}~\bibnamefont{Weinberger}},
  \bibinfo{author}{\bibfnamefont{A.}~\bibnamefont{Schreyer}}, \bibnamefont{and}
  \bibinfo{author}{\bibfnamefont{H.~P.} \bibnamefont{Oepen}},
  \bibinfo{journal}{Phys. Rev. Lett.} \textbf{\bibinfo{volume}{106}},
  \bibinfo{pages}{217207} (\bibinfo{year}{2011}),
  \urlprefix\url{http://link.aps.org/doi/10.1103/PhysRevLett.106.217207}.

\bibitem[{\citenamefont{McGuire and Potter}(1975)}]{McGuire1975}
\bibinfo{author}{\bibfnamefont{T.}~\bibnamefont{McGuire}} \bibnamefont{and}
  \bibinfo{author}{\bibfnamefont{R.}~\bibnamefont{Potter}},
  \bibinfo{journal}{IEEE Trans. Magn.} \textbf{\bibinfo{volume}{11}},
  \bibinfo{pages}{1018Ð38} (\bibinfo{year}{1975}).

\bibitem[{\citenamefont{Chapman et~al.}(2014)\citenamefont{Chapman, Bornstein,
  Ghimire, Mandrus, and Lee}}]{Chapman2014}
\bibinfo{author}{\bibfnamefont{B.~J.} \bibnamefont{Chapman}},
  \bibinfo{author}{\bibfnamefont{A.}~\bibnamefont{Bornstein}},
  \bibinfo{author}{\bibfnamefont{N.~J.} \bibnamefont{Ghimire}},
  \bibinfo{author}{\bibfnamefont{D.}~\bibnamefont{Mandrus}}, \bibnamefont{and}
  \bibinfo{author}{\bibfnamefont{M.}~\bibnamefont{Lee}},
  \bibinfo{journal}{Applied Physics Letters} \textbf{\bibinfo{volume}{105}},
  \bibinfo{pages}{072405} (\bibinfo{year}{2014}).

\bibitem[{\citenamefont{Lee et~al.}(2007)\citenamefont{Lee, Onose, Tokura, and
  Ong}}]{Lee2007}
\bibinfo{author}{\bibfnamefont{M.}~\bibnamefont{Lee}},
  \bibinfo{author}{\bibfnamefont{Y.}~\bibnamefont{Onose}},
  \bibinfo{author}{\bibfnamefont{Y.}~\bibnamefont{Tokura}}, \bibnamefont{and}
  \bibinfo{author}{\bibfnamefont{N.~P.} \bibnamefont{Ong}},
  \bibinfo{journal}{Phys. Rev. B} \textbf{\bibinfo{volume}{75}},
  \bibinfo{pages}{172403} (\bibinfo{year}{2007}),
  \urlprefix\url{http://link.aps.org/doi/10.1103/PhysRevB.75.172403}.

\bibitem[{\citenamefont{Chapman et~al.}(2013)\citenamefont{Chapman,
  Grossnickle, Wolf, and Lee}}]{Chapman2013}
\bibinfo{author}{\bibfnamefont{B.~J.} \bibnamefont{Chapman}},
  \bibinfo{author}{\bibfnamefont{M.~G.} \bibnamefont{Grossnickle}},
  \bibinfo{author}{\bibfnamefont{T.}~\bibnamefont{Wolf}}, \bibnamefont{and}
  \bibinfo{author}{\bibfnamefont{M.}~\bibnamefont{Lee}},
  \bibinfo{journal}{Physical Review B} \textbf{\bibinfo{volume}{88}},
  \bibinfo{pages}{214406} (\bibinfo{year}{2013}).

\bibitem[{\citenamefont{Baibich et~al.}(1988)\citenamefont{Baibich, Broto,
  Fert, Van~Dau, Petroff, Etienne, Creuzet, Friederich, and
  Chazelas}}]{Babibich1988}
\bibinfo{author}{\bibfnamefont{M.~N.} \bibnamefont{Baibich}},
  \bibinfo{author}{\bibfnamefont{J.~M.} \bibnamefont{Broto}},
  \bibinfo{author}{\bibfnamefont{A.}~\bibnamefont{Fert}},
  \bibinfo{author}{\bibfnamefont{F.~N.} \bibnamefont{Van~Dau}},
  \bibinfo{author}{\bibfnamefont{F.}~\bibnamefont{Petroff}},
  \bibinfo{author}{\bibfnamefont{P.}~\bibnamefont{Etienne}},
  \bibinfo{author}{\bibfnamefont{G.}~\bibnamefont{Creuzet}},
  \bibinfo{author}{\bibfnamefont{A.}~\bibnamefont{Friederich}},
  \bibnamefont{and} \bibinfo{author}{\bibfnamefont{J.}~\bibnamefont{Chazelas}},
  \bibinfo{journal}{Phys. Rev. Lett.} \textbf{\bibinfo{volume}{61}},
  \bibinfo{pages}{2472} (\bibinfo{year}{1988}),
  \urlprefix\url{http://link.aps.org/doi/10.1103/PhysRevLett.61.2472}.

\bibitem[{\citenamefont{Ghimire}(2013)}]{GhimireThesis}
\bibinfo{author}{\bibfnamefont{N.~J.} \bibnamefont{Ghimire}}, Ph.D. thesis,
  \bibinfo{school}{University of Tennessee} (\bibinfo{year}{2013}).

\bibitem[{AHE()}]{AHEnote}
\bibinfo{note}{The second term $\rho_{yx} = \mu_0R_HH +\mu_0R_sM$ was modified
  to $\mu_0S_H\rho^2 M\big(1+\frac{\rho_I}{\rho})$, with $\rho_I$ an additional
  fitting parameter, but fits revealed $\frac{\rho_I}{\rho} \ll 1$ over most of
  the $T$ range investigated. Meanwhile, in Fe$_{1/4}$TaS$_2$, $\rho_{xy}$ is
  dictated by the intrinsic AHE at low $T$ and then the extrinsic AHE as $T$
  approaches $T_C$.}

\end{thebibliography}

\end{document}